\documentclass[12pt,a4paper]{article}
\usepackage{hyperref}
\usepackage{appendix}
\usepackage{amsmath}
\begin{document}
\begin{center}

{\LARGE {\bf No accelerating scaling cosmologies\\ \vspace{0.3cm} at string tree level?}}  \\

\vspace{1.5 cm} {\large  Thomas Van Riet }\footnote{ thomas.vanriet @kuleuven.be
}\\
 \vspace{2 cm}
 
 Instituut voor Theoretische Fysica, K.U. Leuven,\\
Celestijnenlaan 200D, B-3001 Leuven, Belgium \\ \vspace{0.3cm}

\vspace{2cm}
\thispagestyle{empty}
{\bf Abstract}
\end{center}

{\small  We investigate the scalar potential in the parametric regime of string moduli space where string loops and higher derivative corrections to 10d supergravity can be ignored and where the fields are rolling down exponential slopes leading to powerlaw FLRW cosmologies, aka scaling solutions. We argue that these scaling solutions, if describing an accelerating expansion, are likely to be perturbatively unstable, for reasons identical to the perturbative instabilities in tree-level dS vacua. 
}

\newpage

\section{Introduction}

It has been known for a while that the asymptotic region of the scalar potentials obtained by compactifying string theory is essentially described by exponential fall offs, in terms of universally present scalar fields, such as the dilaton or volume modulus. Conceptually this can be understood on the basis of general Swampland principles \cite{Ooguri:2018wrx, Hebecker:2018vxz}.  Technically it is a consequence of the fact that, in the proper duality frame, the asymptotic region is expected to be classical: 10d supergravity with branes and orientifold sources. The classical energies of fluxes, curved extra dimensions, and brane/plane sources depend exponentially on the dilaton, volume modulus and potentially many other moduli \cite{VanRiet:2023pnx}. 
As a consequence, the scalar potential $V$ in the classical regime can often be truncated to the following form
\begin{equation} \label{multiexp}
V = \sum_{i=1}^M \Lambda_i \exp\left(\vec{\alpha}_i\cdot\vec{\phi}\,\right)\,,    
\end{equation}
where the universal canonically normalised scalars, such as dilaton and volume, are put in a vector $\vec{\phi}$ with components $\phi^I$, where $I$ runs from $1$ to $N$. There are $M$ exponentials with couplings described by $M$ vectors $\vec{\alpha}_i$ where $i$ runs from $1$ to $M$. The components of these vectors are denoted $\alpha_{iI}$ so that $\alpha$ describes an $M\times N$ real matrix.  The non-universal (ie model dependent) scalars are hidden in the functions $\Lambda_i$ and so these coefficients can carry non-exponential dependence on for instance axion-like scalars. We assume they do not play a relevant role in the asymptotic regions of moduli space. We comment near the end on the strength of our assumptions.

It has become clear in the recent years that it is very unlikely that in the asymptotic region this potential will have any dS critical points \cite{Ooguri:2018wrx, Hebecker:2018vxz, Danielsson:2018ztv, Bena:2023sks, VanRiet:2023pnx, Junghans:2018gdb,Banlaki:2018ayh}. And, in the unlikely case it does, they are most likely perturbatively unstable, consistent with all known classical solutions found thus far, see e.~g.~ \cite{Danielsson:2011au, Andriot:2021rdy}. There is a simple intuition for this \cite{Danielsson:2012et}, crucial to the message of this paper: Circumventing the Maldacena-Nunez nogo theorem \cite{Maldacena:2000mw} requires  the presence of negative tension sources (orientifolds) and so at least one $\Lambda_i$-term will be negative. That term might have a stronger dependence on certain scalar directions (e.~g.~ the O-plane cycle modulus) and so fluctuating the scalars that dominate this term can lower $V$ at the critical point, showing the presence of a tachyon.

Inspired by the difficulties in obtained controlled dS vacua, there is a renewed interest in rolling cosmological scenarios \cite{Agrawal:2018own, Olguin-Trejo:2018zun, Hebecker:2019csg, Cicoli:2020cfj, Cicoli:2021fsd, Cicoli:2021skd, ValeixoBento:2020ujr, Rudelius:2022gbz, Calderon-Infante:2022nxb, Shiu:2023nph, Shiu:2023rxt, Freigang:2023ogu, Cremonini:2023suw,  Hebecker:2023qke} to describe quintessence or inflation.\footnote{See \cite{Cicoli:2023opf, Brandenberger:2023ver} for recent reviews on string cosmology.} And even more recently, the exponential form of the fall-off was the inspiration of \cite{Calderon-Infante:2022nxb,Shiu:2023nph, Shiu:2023rxt} for looking for so-called \emph{scaling} solutions. Scaling solutions, developed for instance in \cite{Halliwell:1986ja, Copeland:1997et}, are powerlaw FLRW solutions:
\begin{equation}
    ds^2 = -dt^2 + t^{2p}[dx^2+dy^2+dz^2]\,,
\end{equation}
where $p$ is the power and when $p>1$ the solution describes an accelerating expansion of the universe. Such solutions are trivially found when sourced by cosmological fluids of constant equation of state, or from scalars with an exponential scalar potential, or a combination of both. This has later been extended to multiple exponentials as in \eqref{multiexp} in various papers such as \cite{Liddle:1998jc, Copeland:1999cs, Collinucci:2004iw, Hartong:2006rt}.

The message of this short note is to use the ideas developed recently in \cite{Shiu:2023nph, Shiu:2023rxt} and the stability criteria developed earlier in \cite{Hartong:2006rt} to demonstrate that accelerating scaling solutions, if they can occur, are most likely going to be perturbatively unstable, and hence share similar problems to dS vacua. Unstable scaling solutions, in the language of dynamical systems, would be repellers, ruling out their practical use since it would require infinite fine-tuning to realise them in the late universe. 

Note that in general \emph{cosmic acceleration is easy to achieve}, despite common believes. This was pointed out long time ago in \cite{Emparan:2003gg}. Just consider for instance string theory compactified on a torus with one type of fluxes. Then the potential has one positive exponential term, due to the positive flux energy. Clearly there is no vacuum, but it is straightforward to contemplate a scenario where the initial velocity of the scalar field points up the exponential slope. After some time the field has to return and near the point of return the scale factor has to obey $\ddot{a}>0$. Yet, in all known models looked at so far the amount of e-folds is very small and not useful for quintessence nor inflation, see eg \cite{Danielsson:2013rza, Marconnet:2022fmx}.\footnote{ It would be very interesting to understand what the general bound is on the number of e-folds in such scenarios.}\footnote{Note that for hyperbolic cosmologies, ($k=-1$) eternal acceleration is also consistent with string theory as first observed in \cite{Andersson:2006du}, see also \cite{Sonner:2006yn, Marconnet:2022fmx}. Although such as scenario is not favoured by observations.} This is in contrast with cosmological scaling solutions that accelerate (with $p>1$). Such solutions have the scalar field rolling down the exponential slope till eternity, at least classically. 

In the following section we briefly recapitulate the essential bits of scaling solutions we need and, in the section after that we present the argument for why accelerating scaling attractors are unlikely to exist. 

\section{Scaling solutions: a quick summary}

Let us first look at the simplest model with a single exponential and let us fix conventions in which $M_p=\sqrt2$, ie, the action is
\begin{equation}
S = \int \sqrt{-g}\left(R -\frac{1}{2}(\partial \phi)^2 - \Lambda \exp(\alpha\phi)\right)\,.     
\end{equation}
The cosmological scaling solution is then
\begin{equation}
a=t^{p} \qquad p =\frac{1}{\alpha^2}\,,\qquad \phi= \frac{-2}{\alpha}\ln(t) + c\,,    
\end{equation}
and is a definitely an attractor (even when including perturbations of a background cosmological fluid) when the solution accelerates $p>1$. Note that V scales like $1/t^2$ and so does the kinetic energy. Hence the name scaling solution: the solution rolls down the potential in such a way that the ratio between the kinetic energy and the potential energy remains constant. Finally, note that \emph{the slow roll} parameter $\epsilon_V$ is given by
\begin{equation}
 \epsilon_V=  \left(\frac{\partial_{\phi} V }{V}\right)^2 = \frac{1}{p}\,.   
\end{equation}
Hence accelerating scaling solutions just require $\epsilon<1$ but not necessarily parametrically small $\epsilon$ and hence slow-roll approximations are not made here.
There can also exist solutions for which also the background barotropic fluid remains a fixed energy ratio with respect to the scalar kinetic and potential energy \cite{Copeland:1997et}. 

The next most involved model is that of so-called \emph{assisted inflation} \cite{Liddle:1998jc}, which despite the name, is often used for late time cosmology:
\begin{equation}
S = \int \sqrt{-g}\left(R -\frac{1}{2}\sum_{I=1}^N(\partial \phi)^2 - \sum_{i=1}^N\Lambda_i \exp(\alpha_i\phi_i)\right)\,.     
\end{equation}
In the notation of the general model (\ref{multiexp}) we are in the situation where $M=N$ and $\alpha$ is a diagonal matrix 
The scaling attractor is then one where all the fields grow like logs and the scale factor is a powerlaw with
\begin{equation}
p=\sum_i \frac{1}{\alpha_i^2}\,.
\end{equation}
The different scalars seem to assist to get more acceleration, hence the name \emph{assisted inflation}. However, this is misleading since it can be shown \cite{Malik:1998gy} that there exists an orthogonal field rotation $\phi\rightarrow \phi'$ such that the kinetic term remains the same but the potential becomes:
\begin{equation}
V = \exp (\alpha_1'\phi_1') U(\phi'_2\ldots, \phi'_N )\,,    \label{newpot}
\end{equation}
where $U$ is a sum of $N$ exponentials depending on $N-1$ scalars: $\phi'_2\ldots, \phi'_N$. The scaling solution is such that $\phi'_1$ is the only active scalar and that the scalars $\phi'_2\ldots, \phi'_N$ are fixed in a critical point of $U$. The field rotation is also such that
\begin{equation}
\left(\frac{1}{\alpha'_1}\right)^2 = \sum_i \frac{1}{\alpha_i^2} = p\,. 
\end{equation}
So we are in the situation of a single exponential really with background fields fixed in a critical point. The stability conditions are the same as the single field one, plus the requirement that the critical point of $U$ is stable, i.~e.~ has a positive Hessian of $U$.

The next best thing is named \emph{generalised assisted inflation} \cite{Copeland:1999cs} and describes the model (\ref{multiexp}) where $M=N$ and the vectors $\vec{\alpha}_i$ are linearly independent. Or, in other words, the matrix $\alpha$ is square and of maximal rank and thus invertible. The scaling attractor is then one where all the fields have log behavior and the power $p$ becomes
\begin{equation}
p =\sum_{ij} A^{-1}_{ij}\,,
\end{equation}
where we sum over the matrix elements of the inverse $A$-matrix, where the $A$-matrix is defined as the matrix of the following inner-products:
\begin{equation}
A_{ij} = \vec{\alpha}_i\cdot \vec{\alpha}_j\,, 
\end{equation}
and is invertible. The stability analysis for this model was carried out in \cite{Hartong:2006rt} and similar to assisted inflation one can perform a field rotation preserving the kinetic term, but which brings the potential to the form (\ref{newpot}). Also here the stability conditions reduce to that of the single field exponential plus the requirement that $U$ has positive Hessian at the fixed point. Crucially for us, one can show \cite{Hartong:2006rt} that this never happens whenever one of the exponentials is negative. This brings us to the next section in which we argue that in the string theory context it implies an obstacle for realising accelerating scaling attractors at asymptotic field values.

\section{Absence of accelerating attractors}
The classical regime of a typical flux compactification is such that it is typically of the form \eqref{multiexp} but with more exponentials than number of scalars: $M>N$. As explained in \cite{Collinucci:2004iw}, the real criterium is whether the rank $R$ of the matrix $\alpha_{iI}$ is maximal ($R=M$) or not $R<M$. The reason is simply that one can perform a field rotation such that only $R$ scalars appear in case $R<N$. It seems that the only way one can now achieve proper scaling solutions is if the vectors $\vec{\alpha}_i$ obey a fine-tuned condition \cite{Collinucci:2004iw}, which can be understood as the condition to rewrite the potential, after field redefinition, again as (\ref{newpot}). 

However, in the real asymptotic regime, things can simplify.\footnote{I am grateful to Flavio Tonioni for explaining this.} For example consider the simplest situation in which $R<M$, namely one scalar ($R=1$), $\phi$ and two exponential terms ($M=2$):
\begin{equation}
V(\phi) = \Lambda_1\exp(\alpha_1\phi) + \Lambda_2\exp(\alpha_2\phi)\,.   
\end{equation}
Imagine, $\alpha_1>\alpha_2>0$, then in the domain of large and negative $\phi$ the second term will dominate over the first and we can approximate
\begin{equation}
V(\phi) \approx \Lambda_2\exp(\alpha_2\phi)\,. 
\end{equation}
This implies that in the asymptotic regimes the rank of the matrix $\alpha$ is maximal and the approximated potential is then of the ``generalised assisted inflation'' \cite{Copeland:1999cs} form and scaling solutions occur. However, in all models where $R=M$ a negative exponential term in the potential implies the accelerating scaling solution is perturbatively unstable as shown in \cite{Hartong:2006rt}. We now argue that without negative terms we cannot have accelerating scaling solutions.

According to extensions of the Maldacena-Nunez nogo theorem \cite{Maldacena:2000mw}, summarized recently in \cite{VanRiet:2023pnx}, de Sitter vacua are only possible with classical ingredients when orientifold planes are allowed since their negative tension is what is required to circumvent the nogo theorems. Hence negative exponentials will occur. A rolling scenario in which for instance the internal space is time-dependent is another way around the MN-like nogo theorems without using orientifolds. Yet this does not lead to dS vacua \cite{Russo:2019fnk} but it can potentially lead to accelerating cosmologies. Indeed, as we made clear earlier, simply having flux on a torus is enough for cosmic acceleration \cite{Emparan:2003gg} but then acceleration is a transient phenomenon with only few e-folds. One can wonder whether accelerating scaling attractors are possible instead? If so, one would have eternal acceleration. Unfortunately, also here the same logic as the Maldacena-Nunez theorem applies. Without orientifold planes one finds that the slow roll parameter is bound, along the lines of \cite{Hertzberg:2007wc, Wrase:2010ew} and it is such that the slow roll parameter for scaling solutions that accelerate $\epsilon=\frac{1}{p}<1$ is not possible.  We prove this now.

We follow the logic pioneered in \cite{Hertzberg:2007wc} and construct the dependence of the various contributions to the scalar potential in 4d on the volume modulus and the string coupling\footnote{See \cite{VanRiet:2023pnx} for a pedagogical introduction to this}. We write the metric in 10d string frame as
\begin{equation}
 ds^2 = \tau^{-2} ds_4^2 + \rho ds^2_6    
\end{equation}
such that the compactification volume is $\rho^3$ as we normalised the volume of $ds_6^2$ to unity (in string units). To get 4d Einstein frame in Planckian units we need to take
$\tau =\rho^{3/2}e^{-\phi}$.

The classical scalar potential with contributions from $p$-form RR-fluxes or 3-form NSNS fluxes ($H_3$) threading the extra dimensions, or from negatively curved\footnote{If positive they would contribute negative terms in 4d.} extra dimensions  and D(3+k) branes\footnote{Again, orientifolds would contribute negative terms in 4d.} wrapping internal $k$-dimensional submanifolds one finds \cite{VanRiet:2023pnx}
\begin{equation}
V = \sum_p \mathcal{U}_p \rho^{3-p} \tau^{-4} + U_{H_3}\rho^{-3}\tau^{-2} + \sum_k \mathcal{U}_k\rho^{\frac{k-6}{2}} \tau^{-3} + \mathcal{U}_R \rho^{-1} \tau^{-2}\,,      
\end{equation}
where we only display the scaling with respect to $\rho$ and $\tau$. Note that this general form of the potential applies to all five string theories, taking into account that only type II string theories allow D-branes and O planes. It is straightforward to add contributions from NS5 branes and they do not change the results below.\footnote{The reason being that their contribution scales as $\rho^{-2}\tau^{-2}$ and we derive below that such scalings do not evade our bounds.} 
The remaining dependencies reside in the various functions labeled $\mathcal{U}$. In terms of canonically normalised scalars $\phi$ (the dilaton) and $\varphi$ (Einstein frame volume) we have
\begin{equation}
\rho = e^{\frac{\phi}{2} +\frac{\varphi}{2\sqrt3}} \,,\qquad \tau = e^{-\frac{\phi}{4} +\frac{\sqrt3\varphi}{4}}  \,.
\end{equation} 
From this and the canonical kinetic term for $\varphi$ and $\phi$ we can compute the kinetic term for $\rho$ and $\tau$ to find
\begin{equation}
    \mathcal{L}_{\text{kinetic}} = -3\frac{(\partial \rho)^2}{\rho^2} - 4\frac{(\partial \tau)^2}{\tau^2}\,.
\end{equation}
We then have the inequality
\begin{equation}
\epsilon_V \geq \frac{1}{3V^2}\left(\rho\frac{\partial V}{\partial \rho}\right)^2 + \frac{1}{4V^2}\left(\tau\frac{\partial V}{\partial \tau}\right)^2    \geq \frac{1}{4V^2}\left(\tau\frac{\partial V}{\partial \tau}\right)^2\,.
\end{equation}
To bound this further we introduce some shorthand notation. Denote each term in the potential with $V_i$ and the negative power of $\tau$ as $a_i$. We see that $a_i$ can only be $-2, -3, -4$. We then have
\begin{equation}
\epsilon \geq \frac{\left( \sum_i \tfrac{|a_i|}{2} V_i\right)^2}{\left(\sum_i V_i\right)^2} \geq 1\,,
\end{equation}
since all $a_i$ are negative, all $V_i$ positive and $\frac{|a_i|}{2}\geq 1$.

To summarize:  any asymptotic regime which allows scaling solutions that are stable must not have negative terms as proven in \cite{Hartong:2006rt}. This means that O-plane tensions must be subdominant. But then the slow roll parameter can never be small enough to allow for cosmic scaling solutions that accelerate. Transient acceleration is however not excluded.

\section{Outlook}
The difficulty of finding dS vacua in controlled regimes of string compactifications \cite{Danielsson:2018ztv, Bena:2023sks, VanRiet:2023pnx} has revived the interest in finding quintessence like scenarios in which scalar fields roll down the potential. Given that the potential slopes in the controlled regimes are exponential one expects these rolling scenarios to correspond to so-called scaling solutions \cite{Copeland:1997et} which exhibit powerlaw behavior for the scale factor. Using the stability results of \cite{Hartong:2006rt} together with properties of the string effective action in the controlled regime (detailed in the Appendix) we have argued that it is equally problematic to find such accelerating scaling attractors as it is to find meta-stable dS vacua. 

The strongest assumption made here is that we are in an asymptotic regime where we can justify that a higher exponential power is subdominant over the smaller: eg $e^{3\phi}$ can be ignored of there is a term that scales as $e^{2\phi}$. This is typically true in asymptotic regimes where every term used in the potential contributes roughly equal amounts. This approximation implied that we can always restrict our potential to $R$ exponential terms where $R$ is the rank of the matrix $\alpha$. Such a truncation can be proven to be exact in terms of finding scaling solutions that can be seen as fixed points of an autonomous system viewpoint of the cosmological dynamics \cite{Collinucci:2004iw, Hartong:2006rt} and can be understood even beyond perturbative fixed point analysis \cite{Shiu:2023nph, Shiu:2023rxt}. 

However, this approximation can fail if some moduli are not at asymptotic values. Not all moduli should be at asymptotic values to have parametric perturbative control, just imagine a modulus that is measuring the relative size of two cycles. Then we can be in a situation where $R<M$ with negative terms in the function $U$, without it creating instabilities. Consider for example the following situation where $R=2$ and $M=3$
 \begin{equation}
     V = e^{\frac{\phi}{\sqrt{p}}}\left( \Lambda_1 e^{a\varphi} +\Lambda_2e^{b\varphi} +\Lambda_3 e^{c\varphi}\right)\,.
 \end{equation}
with $\Lambda_2<0$ and $\Lambda_{1,3}>0$. It is easy to find values for $a, b, c$ such that there is a stable minimum of the function in between brackets (the $U$-function). But this is similar, in fact identical, to the discussion of finding meta-stable dS at tree-level. Then the $U$-function above plays the role of the full scalar potential and the negative term (the one with $\Lambda_2$) typically induces tachyons \cite{Danielsson:2012et, Junghans:2016uvg}. The intuition for that is rather straightforward but cannot (yet?) be turned into a nogo theorem: the negative term corresponds to orientifold tension and, of all terms, is the one that most strongly varies as the volume of the orientifold cycle varies. Hence fluctuations of that volume induce instabilities as they can lower the energy.

Another approximation we have relied on is the rather strict exponential form of the potential and a canonical kinetic term for the scalars involved in the cosmological evolution. Obviously the kinetic terms in EFTs derived from compactification are almost never Riemann-flat but have Riemann-flat subspaces. A typical example is an axion coupling to the scalars $\phi^I$ which leads to hyperbolic kinetic terms and can lead to new scaling cosmologies, see eg
\cite{Sonner:2006yn, Brinkmann:2022oxy}. But axion-momentum only slows down cosmic expansion so our conclusions still hold. Regarding the exponential form of the scalar potential, we only require it to have the form \eqref{newpot} where $U$ is a function with a positive minumum, it by itself does not have to be a sum of exponentials. 

One can also wonder to what extend assymptotic regimes correspond to classical regimes at the two-derivative level. For instance reference \cite{Fraiman:2023cpa} considers a regime where there seems to be no such picture yet the scalar potential possesses the exponential features and perturbative instabilities seem to be hard to avoid. This is encouraging and supports our claims. 

Finally, the fact that we find the form \eqref{newpot} is reminiscent of the claim made in \cite{Hebecker:2023qke} that asymptotic acceleration is similar to having higher-dimensional dS vacua. Just imagine a higher-dimensional de Sitter vacuum. This means we have a theory in $4+d$ dimensions with a scalar potential $U(\phi^I)$ that has a local minimum at a positive value. Now one can dimensionally reduce that dS solution by assuming that some spatial directions are circles. Then the scalar potential in 4d is
\begin{equation}
V(\phi^I, \varphi) = e^{\frac{\varphi}{\sqrt{p}}}U(\phi^I) \,,   
\end{equation}
with $\varphi$ the canonically normalised volume modulus of the $d$-dimensional torus (inside the planar dS Poincar\'e patch in $4+d$ dimensions) with \cite{Rosseel:2006fs}
\begin{equation}
    p = \frac{d+2}{d} >1\,.
\end{equation}
In other words, higher dimensional de Sitter vacua reduce to 4-dimensional accelerating scaling attractors with the function $U$ in \eqref{newpot} playing the role of the higher-dimensional scalar potential, with its expected instabilities near the boundary of moduli space. 

\section*{Acknowledgments}
I benefited from discussions with Gerben Venken, Gary Shiu and Flavio Tonioni. I furthermore thank Flavio Tonioni and Gary Shiu for useful feedback on an earlier draft.


\bibliographystyle{utphys-modified}
\small{
\bibliography{refs}}
\end{document}